\documentclass[preprint,showpacs,amsmath,amssymb,amsfonts,prl,aps]{revtex4}
\usepackage{amsmath}
\usepackage{amssymb}
\usepackage{graphicx}
\usepackage{color}
\usepackage{float}
\usepackage[dvipsnames]{xcolor}

\begin{document}

\title{Source-free exchange-correlation magnetic fields in density functional theory}
\author{S. Sharma$^1$}
\email{sharma@mpi-halle.mpg.de}
\author{E. K. U. Gross$^1$}
\author{A. Sanna$^1$}
\author{J. K. Dewhurst$^1$}
\affiliation{$^1$Max-Planck Institut f\"ur Microstrukture Physics, Weinberg 2, D-06120 Halle, Germany}

\date{\today}

\begin{abstract}
Spin-dependent exchange-correlation energy functionals in use today
depend on the charge density and the magnetization density:
$E_{\rm xc}[\rho,{\bf m}]$.
However, it is also correct to define the functional in terms of
the curl of ${\bf m}$ for physical external fields:
$E_{\rm xc}[\rho,\nabla\times{\bf m}]$. The exchange-correlation
magnetic field, ${\bf B}_{\rm xc}$, then becomes source-free. We study this
variation of the theory by uniquely removing the source term from
local and generalized gradient approximations to the functional.
By doing so, the total Kohn-Sham moments are improved for a wide range of
materials for both functionals. Significantly, the moments for the pnictides
are now in good agreement with experiment.
We also predict dramatic differences in the spatial
geometry of ${\bf B}_{\rm xc}$ for the pnictides.
Our source-free method is simple to implement in all existing density
functional theory codes.
\end{abstract}

\maketitle

\begin{figure}[H]
\centerline{\begin{tabular}{c}
\includegraphics[width=0.75\columnwidth, clip]{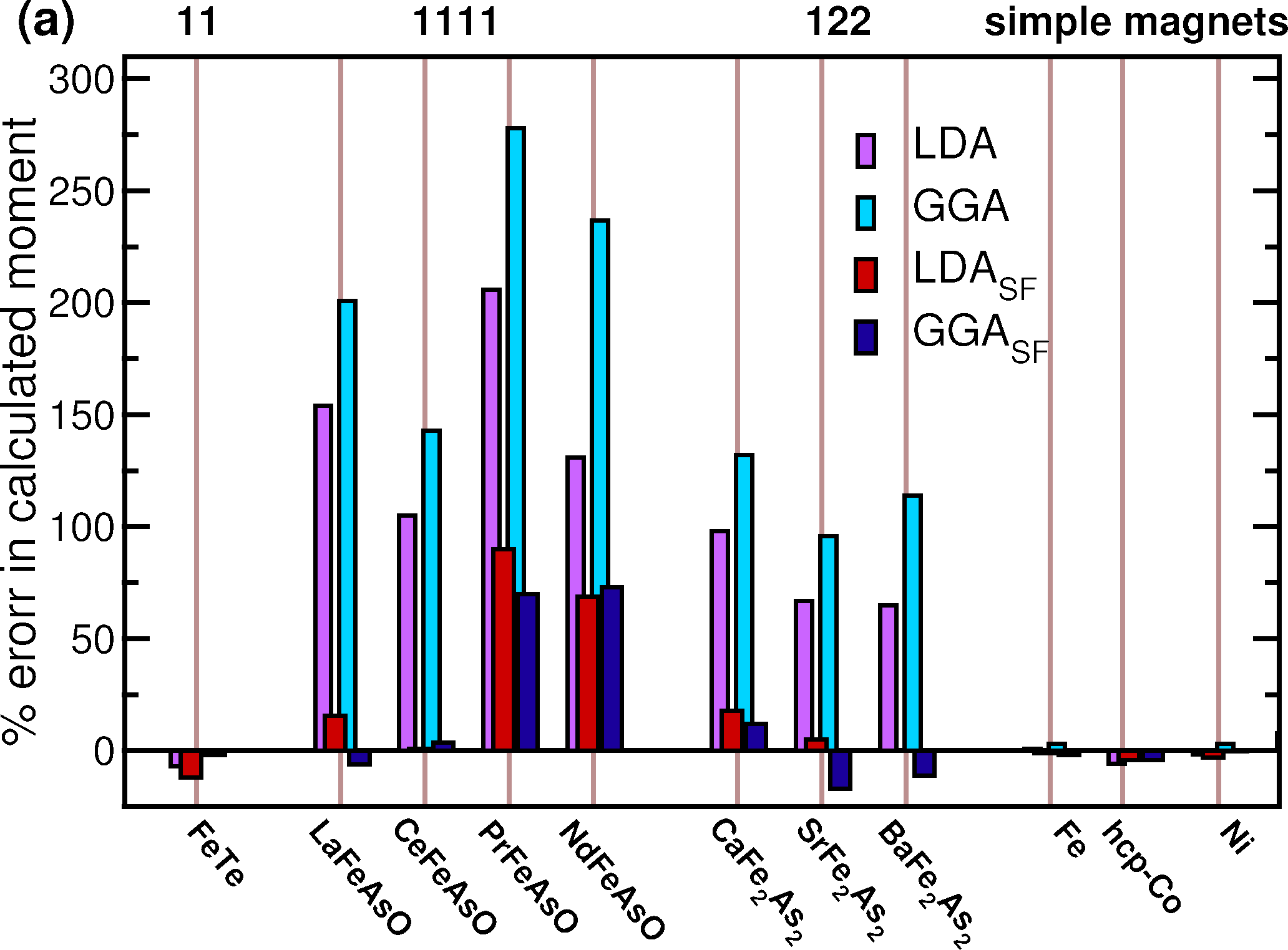} \\
\includegraphics[width=0.75\columnwidth, clip]{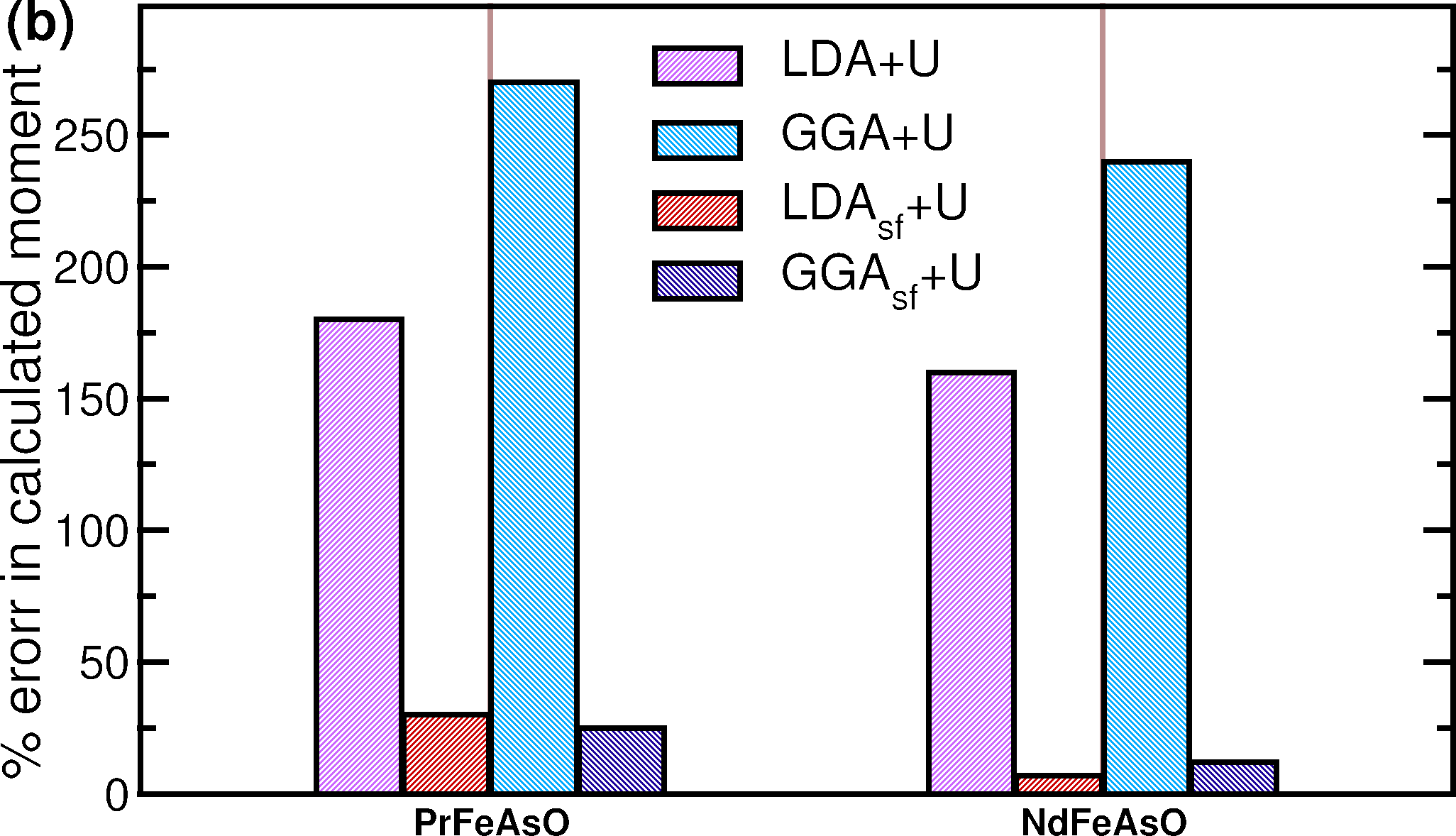} 
\end{tabular}}
\caption{\label{perr} Percentage deviation of the calculated magnetic moment from
 experimental data for 11, 1111, 122 pnictides as well as elemental solids. 
 (a) Results calculated using LSDA (pink), GGA (cyan), and their source-free counterparts 
 LSDA$_{\rm SF}$ (red) and GGA$_{\rm SF}$ (blue) (b) same as (a) but by adding an on-site Coulomb
 repulsion $U$ on the $f$-states. Root-mean-square-percentage errors are
 LSDA: 90.2\%, GGA: 143\%, LSDA$_{\rm SF}$: 34\%, GGA$_{\rm SF}$: 31\%,
 LSDA$_{\rm SF}$+U: 16\%, GGA$_{\rm SF}$+U: 11\%.}
\end{figure}

Density functional theory (DFT)\cite{hk,ks} has proven enormously successful
for calculating
the electronic structure of both molecules and solids. Lattice structures,
phonon spectra and many other properties are now routinely calculated.
Magnetism presents more of a mixed picture.
Simple magnets, such as elemental solids (Fe, Co and Ni), are
well-described by the local spin density
approximation (LSDA)\cite{lda} or the
generalized gradient approximations (GGA)\cite{pbe}, at least as far as
total moments are concerned.
However, both LSDA and GGA perform poorly for
the iron pnictide and related materials\cite{dai-rev,mazin-rev,Kordyuk_Review_IronSC,Dai_FeSCMagnetism_NatPhys_2012}
for which they greatly overestimate the moments by factors of two or more (see Fig.~\ref{perr}).
This has been an impediment to investigating the microscopic magnetic structure\cite{Yildirim_LaOFeAs_PRL2008,Sing_LaOFeAs_PRL2008,Ma_LaOFeAs_PRB2008,our-ce,Ortenzi_scaling_PRL2015}, related response functions\cite{Essenberger_chiFeSe_PRB2012} and superconductivity~\cite{Mazin_LaOFeAs_PRL2008,Lischner_chiFeSe_2015,Essenberger_SC_FeSe_PRB2016} of these materials with density functional methods.

Most approximate spin-dependent
exchange-correlation energy functionals, $E_{\rm xc}[\rho,{\bf m}]$, use the density and
magnetization as their arguments\cite{vonBarth1972,sandratskii86,kubler88,nikitas}. 
This form is dictated by the many-body Hamiltonian used originally in the context of spin DFT (SDFT) by
von Barth and Hedin\cite{vonBarth1972}:
\begin{align}\label{hmlvbh}
 \hat{H}=\sum_{i=1}
 -\frac{1}{2}\nabla_i^2+V_{\rm ext}({\bf r}_i)
 +\vec{\sigma}\cdot{\bf B}_{\rm ext}({\bf r}_i)
 +\frac{1}{2}\sum_{j\ne i}\frac{1}{\left|{\bf r}_i-{\bf r}_j\right|},
\end{align}
where $V_{\rm ext}$ and ${\bf B}_{\rm ext}$ are the external scalar potential
and magnetic fields, respectively; and the sum runs to the number of particles.
The external magnetic field was assumed to be an unconstrained vector
field in the original formulation of SDFT.
Physical magnetic fields are not unconstrained but rather
the curl of a vector potential,
i.e. ${\bf B}_{\rm ext}=\nabla\times {\bf A}_{\rm ext}$.
With this constraint it is possible to demonstrate
(see \cite{eschrig,capelle} and Appendix) that the
exchange-correlation functional can be chosen to depend on
the spin current
$\nabla\times {\bf m}({\bf r})$ instead of ${\bf m}({\bf r})$:
$\widetilde{E}_{\rm xc}[\rho,\nabla\times{\bf m}]$.

An immediate consequence of this is that the functional derivative of
$\widetilde{E}_{\rm xc}[\rho,\nabla\times{\bf m}]$ with respect to
${\bf m}({\bf r})$ is of the form
$\widetilde{\bf B}_{\rm xc}({\bf r})
 \equiv\delta \widetilde{E}_{\rm xc}/\delta{\bf m}({\bf r})
 =\nabla\times {\bf A}_{\rm xc}({\bf r})$ which
implies $\nabla\cdot{\bf B}_{\rm xc}({\bf r})=0$. In other words,
the exchange-correlation magnetic field is {\em source-free}.

An essential aspect of this version of SDFT, and one which has been
seemingly overlooked in the past, is that the Kohn-Sham magnetization
${\bf m}({\bf r})$
obtained from the exact $\widetilde{\bf B}_{\rm xc}$ is {\em not} itself exact,
but rather only its curl is.
The difference between the two is a curl-free function which is therefore the
gradient of some scalar function: $\nabla f({\bf r})$.
Thus one loses some information
about the magnetization by using the source-free theory but not, as it turns
out, the total moment. For finite systems, the
total moment obtained from the Kohn-Sham magnetization
using $\widetilde{\bf B}_{\rm xc}$ is also exact because the integral
of $\nabla f({\bf r})$ over all space is zero.
This is not true for periodic boundary conditions, however.
In this case, the functional domain has to be augmented with the total
moment vector ${\bf M}$, thus
$\widetilde{E}_{\rm xc}\equiv \widetilde{E}_{\rm xc}[\rho,\nabla\times{\bf m},{\bf M}]$.
This is analogous
to the macroscopic polarization required as an extra variable in presence of an
external electric field applied to a solid\cite{ggg}.
The variable conjugate to ${\bf M}$ now has to be included in the calculation;
this variable is clearly a constant magnetic field and corresponds to an
${\bf A}_{\rm ext}$ which diverges at large distance. 

Functionals in common use, such as LSDA and GGA, are not, in general,
source-free. This is confirmed simply by computing
$\nabla\cdot{\bf B}_{\rm xc}({\bf r})$ for a magnetic material.
One may therefore reasonably ask: how can LSDA or GGA be modified
so that they do have this property,
i.e. how can any approximate ${\bf B}_{\rm xc}$ be made source-free?
We appeal to Helmholtz's theorem which states that any vector field
on a domain in $\mathbb{R}^3$, which is twice differentiable can
be decomposed into a curl-free component and a source-free component.
This decomposition is unique for given boundary conditions.
Thus let $\phi$ be the solution to Poisson's equation (in atomic units)
\begin{align}
 \nabla^2\phi({\bf r})=-4\pi\nabla\cdot{\bf B}_{\rm xc}({\bf r})
\end{align}
and define
\begin{align}
 \widetilde{\bf B}_{\rm xc}({\bf r})\equiv{\bf B}_{\rm xc}({\bf r})
 +\frac{1}{4\pi}\nabla\phi({\bf r})
\end{align}
then $\nabla\cdot\widetilde{\bf B}_{\rm xc}({\bf r})=0$,
i.e. $\widetilde{\bf B}_{\rm xc}$ is source-free.
It is important to note that that the scalar part of the potential,
$V_{\rm xc}({\bf r})$, is not directly affected by this procedure.
This modified functional has certain intrinsic properties:
(a) it is still correct for homogeneous electron gas (HEG) because
${\bf B}^{\rm HEG}_{\rm xc}$ is a constant implying that
$\nabla\cdot{\bf B}_{\rm xc}({\bf r})=0$ and therefore this
modification has no effect, (b) since $\widetilde{\bf B}_{\rm xc}$ is obtained
by solving Poisson's equation, the functional is intrinsically non-local,
in other words, the field at ${\bf r}$ depends on the magnetization everywhere,
(c) $\widetilde{\bf B}_{\rm xc}$ is necessarily non-collinear,
(d) ${\bf m}({\bf r})\times\widetilde{\bf B}_{\rm xc}({\bf r})$
is non-zero and hence
will contribute to spin-dynamics even in the absence of the external field\cite{our-exx},
(e) $\widetilde{\bf B}_{\rm xc}$ is not, in general, the functional derivative of
an energy functional (see Appendix),
(f) the procedure is simple to implement in any code since all codes have a
Poisson equation solver, and (g) very little computational effort is
needed for the modification.

We perform an additional modification of the functional which effectively
enhances the spin splitting.  
It comprises of a simple scaling of the input magnetization
$E_{\rm xc}[\rho,{\bf m}]\rightarrow E_{\rm xc}[\rho,s{\bf m}]$ and then a
further scaling of the resultant magnetic field
${\bf B}_{\rm xc}\rightarrow s{\bf B}_{\rm xc}$ in order to keep the functional
variational with respect to ${\bf m}$. This scaling is reminiscent of that
performed by Ortenzi {\it el al.}\cite{ortenzi}.
The scaling factor $s$ is chosen
empirically and we find that $s=1.12$ and $s=1.14$ are good choices for
LSDA and GGA respectively, for a diverse set of materials. We note that this factor,
though empirical, is not a material-dependent parameter.

In the present work we apply this procedure to LSDA and GGA:
first we enhance the strength of the exchange
splitting and then modify the functional
in a unique way to become source-free. 
These new source-free functionals have the effect that:
(a) the good moments of LSDA for elemental solids are retained, 
(b) the large overestimation of pnictide moments is cured, and 
(c) the GGA now yields better results than LSDA for both classes of materials.
These results are illustrated in Fig. \ref{perr}, where we show percentage
errors for both common and pnictide materials, with both LSDA and GGA and
their source-free counterparts. Our implementation is publicly available in the
Elk Code\cite{elk} and so can be applied to many more such materials.


\begin{table}[tbh]
\caption{Magnetic moment (in $\mu_{\rm B}$) per Fe atom for pnictides and per
 magnetic atom for the rest of the materials. For pnictides the moment is known
 to be highly sensitive to the structural details and hence the references for
 experimental structural data are cited in the first column and for magnetic
 moment in the second column.}\label{table1}
\begin{tabular}{c|c|c|c|c|c}
Material                    & Expt               & LSDA & LSDA$_{\rm SF}$ & GGA  & GGA$_{\rm SF}$     \\ \hline\hline 
LaFeAsO\cite{La1111s}       & 0.63\cite{La1111m} & 1.60 & 0.73            & 1.92 & 0.59                 \\ 
LaFeAsO\cite{ggao}          &                    & 1.39 & 0.7             & 1.8  & 0.58                  \\ \hline       
CeFeAsO\cite{Ce1111s}       & 0.8\cite{Ce1111m}  & 1.64 & 0.81            & 1.95 & 0.83               \\ \hline
PrFeAsO\cite{Pr1111s}       & 0.5\cite{Pr1111m}  & 1.53 & 0.99            & 1.89 & 0.85               \\ \hline
NdFeAsO\cite{Nd1111s}       & 0.54\cite{Nd1111m} & 1.24 & 0.91            & 1.82 & 0.93               \\ \hline
CaFe$_2$As$_2$\cite{Ca122s} & 0.8\cite{Ca122m}   & 1.59 & 0.95            & 1.86 & 0.90               \\ \hline
SrFe$_2$As$_2$\cite{Sr122s} & 0.94\cite{Sr122m}  & 1.57 & 0.98            & 1.84 & 0.78               \\ \hline
BaFe$_2$As$_2$\cite{Ba122s} & 0.87\cite{Ba122m}  & 1.43 & 0.87            & 1.84 & 0.78               \\
BaFe$_2$As$_2$\cite{ggao}   &                    & 1.38 & 0.73            & 1.67 & 0.59                \\ \hline
FeTe\cite{FeTes}            & 2.25\cite{FeTem}   & 2.10 & 1.73            & 2.25 & 1.85               \\ \hline
bcc-Fe                      & 2.2                & 2.15 & 2.22            & 2.27 & 2.16               \\ \hline
hcp-Co                      & 1.7                & 1.63 & 1.60            & 1.67 & 1.61               \\ \hline
Ni                          & 0.65               & 0.64 & 0.63            & 0.67 & 0.65               \\ \hline
Ni$_3$Al                    & 0.077              & 0.17 & 0.1225          & 0.1825 & 0.1725           \\ \hline
ZrZn$_2$                    & 0.085              & 0.21 & 0.197           & 0.283  & 0.257            \\ \hline
\end{tabular}
\end{table}

The percentage deviation in the magnetic moment from experiment are presented
in Fig. \ref{perr}(a). We note that the magnetic moments calculated using
LSDA or GGA for simple magnets (Fe, Co and Ni) are already in very good
agreement with experiment with maximum deviation of 8\%. This is in contrast
the moments for pnictides which deviate strongly from experiment with a
maximum error of 278\%. The moments calculated using source-free LSDA and GGA are
also presented in Fig. \ref{perr}(a) and for simple magnets they are of
the same quality as that of the unmodified functionals.
The fact that the integrated moments are the same does not
necessarily imply that the magnetization densities are similar at each point in space.
However, we find that for the simple magnets the two densities are fairly close at each point in space.
In the case of pnictides, the moments show dramatic improvement.
At a first glance it appears that
LSDA$_{\rm SF}$/GGA$_{\rm SF}$ substantially reduces the Fe moment for all
pnictides compared to the corresponding LSDA/GGA value. A closer inspection, however, 
reveals that this reduction is, as it should be, highly selective in that the moment 
in SrFe$_2$As$_2$ is reduced by $\sim$30\% while on LaFeASO by $\sim$60\% compared to 
the LSDA/GGA results. 
The maximum deviation is now less than 25\% for all materials with
the exception of NdFeAsO and PrFeAsO (see Table \ref{table1}).

\begin{table}[tbh]
\caption{Magnetic moment (in $\mu_{\rm B}$) per atom. Calculations are
 performed using LSDA$+U$, GGA$+U$, LSDA$_{\rm SF}+U$ and GGA$_{\rm SF}+U$. 
 The values of $U$ and $J$ used for these calculations are given
 in Ref. \onlinecite{uj}.}\label{table2}
\begin{tabular}{c|c|c|c|c|c}
Material & Expt      & LSDA     & GGA       & LSDA$_{\rm SF}$   & GGA$_{\rm SF}$    \\ \hline\hline
PrFeAsO  & Fe: 0.5   & 1.40     & 1.9       & 0.65              & 0.63               \\ 
         & Pr: 0.87  & 0.30     & 0.30      & 0.81              & 0.83              \\ \hline
NdFeAsO  & Fe: 0.54  & 1.42     & 1.84      & 0.50              & 0.61              \\
         & Nd: 0.9   & 2.44     & 1.25      & 0.80              & 0.89               \\ \hline
\end{tabular}
\end{table}

For NdFeAsO and PrFeAsO the source-free functionals provide a
considerable improvement over unmodified LSDA or GGA, but the
percentage deviation from experiment is still relatively large.
We find the reason behind this to be the moment on the rare earth atoms.
In these materials the moment of the rare-earth atom is known to be strongly 
coupled to the moment on the Fe atoms\cite{Nd1111m} and the 
rare earth moment is not accurately described by the source-free functional
alone\cite{mwu}.
In order to treat these we use the well-established
method\cite{ldapu} of applying an on-site Coulomb repulsion $U$. 
It is important to mention that $U$ was applied only
to the $f$-states of the rare-earth atom and chosen to reproduce the
experimental moment of that atom only. Nevertheless, this substantially improves
the  moment on the Fe sites (see Table \ref{table2}).
Like experiments we find
the moment on rare-earths to be in-plane and oriented perpendicular to the moment 
on the Fe atoms. The effect of the source-free functional is particularly
apparent in these cases since without it the correct
moment on either atom cannot be obtained for any choice of $U$.

A material that also requires special attention is LaFeAsO, perhaps the most
studied pnictide of all. Reported experimental values of
the magnetic moment range from 0.36$\mu_{\rm B}$\cite{lafeaso0.36} to
0.8$\mu_{\rm B}$\cite{lafeaso0.8} making it difficult to know to what
our theoretical results should be compared. Perhaps the best choice is
with the more recent experimental value which lies in between these two
extremes, 0.63$\mu_{\rm B}$\cite{La1111m}. This experiment was performed at
low temperature (2K) which is closest to our theoretical ideal of zero temperature.
Another reasonable choice would be 0.8$\mu_{\rm B}$\cite{lafeaso0.8} since, like
our theoretical work, these experiments are performed on single crystals.
In either case, our results with source-free functionals still show a maximum
deviation of 25\%.  

Note that the two steps which comprise this method 
(i.e. scaling and making the functional source-free) must be performed in combination;
each applied alone yields unreasonable results (see Table \ref{tab_sm} in Appendix).

A means of summarizing the overall quality of results is the
root-mean-square percentage error (RMSPE) which one expects to be
reduced on improving the functional, by going, for example, from LSDA to the
more sophisticated
GGA. To the contrary, the value of RMSPE for LSDA is 90.2\% and for
GGA is 143\%, i.e. the
quality of results deteriorates on improving the functional by adding gradients. 
These errors are greatly reduced to 34.3\% with LSDA$_{\rm SF}$ and
30.6\% with GGA$_{\rm SF}$. Furthermore, once the LSDA$_{\rm SF}+U$ and
GGA$_{\rm SF}+U$ results are considered the RMSPE are 16\% for source-free LSDA
and 11\% for source-free GGA, indicating that removal of the source term results
in GGA$_{\rm SF}$ performing better than LSDA$_{\rm SF}$.

It is also appropriate to ask under which circumstances do these
source-free functionals fail?
For the class of materials which are well known
to be close to quantum critical points and for which spin-fluctuations dominate
the physics\cite{ni3al,znzr}, i.e. Ni$_3$Al and ZnZr$_2$, the functional fails in that
percentage errors compared to experiments, despite being an improvement
over LSDA/GGA, are still large.
Another aspect where LSDA$_{\rm SF}$ and GGA$_{\rm SF}$ are deficient is the fact
that they are potential functionals and not energy functionals. Thus one
cannot compare the total energies of various structural and magnetic configurations.
Hence in the present work we have used
experimental crystal structures. However, PBE is well known to be excellent for
structural optimization in most cases, including magnetic pnictides, with lattice
parameters very close to experiments and small deviations in the
position of As atom~\cite{Pickett_PRL2008,Yildirim_PRL2009,Colonna_CaFe2As2-BaFe2As2_PRB2011,Mazin_ProblemsDFT_PRB2008,our-ce}.
Hence one could perform a full optimization
using PBE followed by a calculation using LSDA$_{\rm SF}$ or GGA$_{\rm SF}$.
To test this we have performed such calculations for BaFe$_2$As$_2$ and LaFeAsO
and find that the moment obtained using LSDA and GGA are still highly
over-estimated (maximum error of 180\%), while the moments obtained using 
LSDA$_{\rm SF}$ and GGA$_{\rm SF}$ are again within 25\% of experiment (see Table I).

\begin{figure}
\centerline{
\includegraphics[width=0.75\columnwidth, clip]{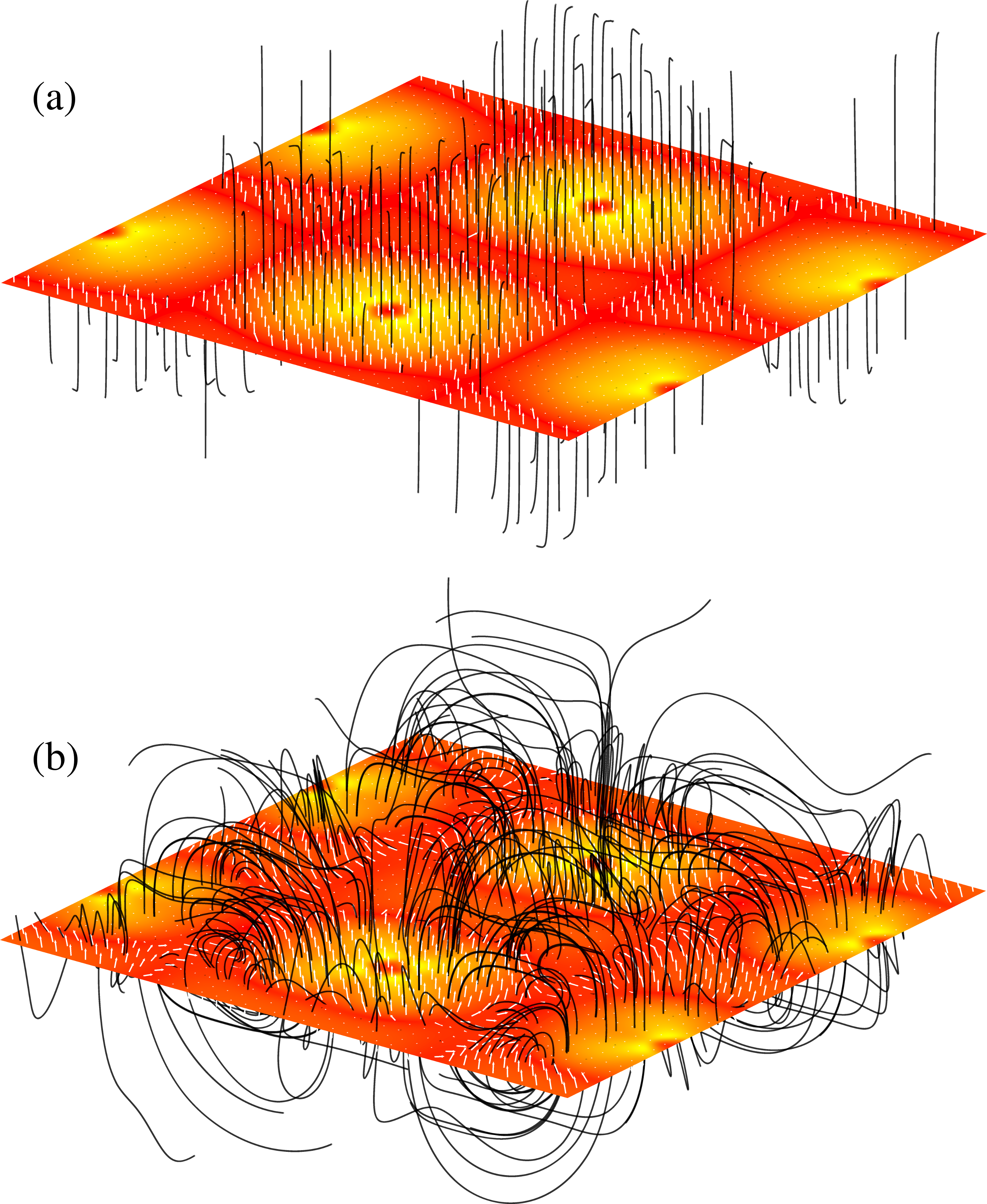} }
\caption{\label{figbxc}The vector field ${\bf B}_{\rm xc}$ for
 BaFe$_2$As$_2$.
 Plot (a) is LSDA and plot (b) is source-free LSDA. The colored plane
 contains the Fe atoms and
 shows the magnitude of ${\bf B}_{\rm xc}$, with the small white arrows
 indicating direction.
 The black field lines originate from a regular grid
 in the plane and follow the vector field.}
\end{figure}
It is enlightening to see how ${\bf B}_{\rm xc}$ of LSDA and GGA
and their new source-free versions differ spatially. We plot this for the
case of BaFe$_2$As$_2$ in Fig. \ref{figbxc}. The field lines for LSDA
are unphysical in the sense that they begin and end at different points
whereas the source-free field lines are always closed. This means that they have to follow
more complicated paths in the crystal, a fact evident from Fig. \ref{figbxc}.

To summarize: motivated by an exact property of spin current DFT, we
removed the source term from the ${\bf B}_{\rm xc}$ of
LSDA and GGA. The spin splitting was also enhanced by a simple scaling of the
input magnetization and output field. The resulting functionals were found
to produce moments which were in better agreement with experiment.
This improvement was particularly pronounced for the pnictides where errors
were reduced from 100-200\% down to 25\% or less. 
We hope that our findings will spur the development of exchange-correlation
energy functionals whose resultant magnetic fields are manifestly
source-free.

\section{Acknowledgments}
The authors would like to thank Kieron Burke and Carsten Ullrich for useful
discussions.

\section{Appendix}
\subsection{Showing ${\bf B}_{\rm xc}$ can be chosen source-free}
Let
\begin{align}
 E[V_{\rm ext},B_{\rm ext}]=\min_{|\Psi\rangle}\langle\Psi|\hat{H}|\Psi\rangle,
\end{align}
where $\hat{H}$ is given in Eq. (1) and the minimization is
over all $N$-electron states $|\Psi\rangle$,
be the total energy as a functional of the external potential
and magnetic field. This can be written as a constrained minimization
\begin{align}
 E[V_{\rm ext},B_{\rm ext}]&=\min_{(\rho,{\bf m})}
 \min_{|\Psi\rangle\rightarrow(\rho,{\bf m})}\langle\Psi|\hat{H}|\Psi\rangle \\
 &=\min_{(\rho,{\bf m})}\left\{\int d^3r\,V_{\rm ext}({\bf r})\rho({\bf r})
  +\int d^3r\,{\bf B}_{\rm ext}({\bf r})\cdot{\bf m}({\bf r})
  +F[\rho,{\bf m}]\right\},
\end{align}
where
\begin{align}
 F[\rho,{\bf m}]\equiv\min_{|\Psi\rangle\rightarrow(\rho,{\bf m})}
 \langle\Psi|\hat{T}+\hat{V}_{\rm ee}|\Psi\rangle
\end{align}
is a universal functional of density and magnetization; $\hat{T}$ and
$\hat{V}_{\rm ee}$ are the kinetic and electron-electron interaction parts
of the Hamiltonian, respectively.
Likewise, the non-interacting kinetic energy functional is defined as
\begin{align}
 T_s[\rho,{\bf m}]\equiv\min_{|\Psi\rangle\rightarrow(\rho,{\bf m})}
 \langle\Psi|\hat{T}|\Psi\rangle
\end{align}
from which is obtained the exchange-correlation energy functional
$E_{\rm xc}[\rho,{\bf m}]\equiv F[\rho,{\bf m}]-T_s[\rho,{\bf m}]-E_{\rm H}[\rho]$,
where $E_{\rm H}$ is the usual Hartree energy.

However, if we assume that the external magnetic field is physical, i.e.
${\bf B}_{\rm ext}({\bf r})=\nabla\times{\bf A}_{\rm ext}({\bf r})$,
and the magnetization tends to zero at large distance,
then the classical energy of the external magnetic field can be written as
\begin{align}\label{engyext}
 \int d^3r\,\left(\nabla\times{\bf A}_{\rm ext}({\bf r})\right)\cdot{\bf m}({\bf r})
 =\int d^3r\,{\bf A}_{\rm ext}({\bf r})
  \cdot\left(\nabla\times{\bf m}({\bf r})\right).
\end{align}
Thus we can define another universal functional $\widetilde{F}[\rho,\nabla\times{\bf m}]$
as
\begin{align}
 \widetilde{F}[\rho,\nabla\times{\bf m}]
 \equiv\min_{|\Psi\rangle\rightarrow(\rho,\nabla\times{\bf m})}
 \langle\Psi|\hat{T}+\hat{V}_{\rm ee}|\Psi\rangle,
\end{align}
with analogous functionals $\widetilde{T}[\rho,\nabla\times{\bf m}]$ and
$\widetilde{E}_{\rm xc}[\rho,\nabla\times{\bf m}]$.
On the space of densities obtained from physical external magnetic fields
we have that
$E_{\rm xc}[\rho,{\bf m}]=\widetilde{E}_{\rm xc}[\rho,\nabla\times{\bf m}]$.
This implies that the total energy obtained from both functionals is
also the same for physical densities.

The equality of the functionals does not hold in general for densities obtained
from external magnetic fields which have a source term.
A consequence of this is that the unconstrained functional derivative of
$E_{\rm xc}$ with respect to ${\bf m}({\bf r})$ is different for the two
functionals, i.e. ${\bf B}_{\rm xc}({\bf r})\ne\widetilde{\bf B}_{\rm xc}({\bf r})$ in
general. The functional derivative of $\widetilde{E}_{\rm xc}$ can be further
evaluated as
\begin{align}
\begin{split}
 \widetilde{\bf B}_{\rm xc}({\bf r})&\equiv
 \left.\frac{\delta \widetilde{E}_{\rm xc}[\rho,\nabla\times{\bf m}]}
  {\delta{\bf m}({\bf r})}\right|_{\rho} \\
 &=\int d^3r'\,
 \frac{\delta\left(\nabla\times{\bf m}({\bf r}')\right)}{\delta {\bf m}({\bf r})}
 \left.\frac{\delta \widetilde{E}_{\rm xc}[\rho,\nabla\times{\bf m}]}
 {\delta\left(\nabla\times{\bf m}({\bf r}')\right)}\right|_{\rho} \\
 &=\nabla\times{\bf A}_{\rm xc}({\bf r}),
\end{split}
\end{align}
proving that $\widetilde{\bf B}_{\rm xc}$ is indeed source-free.
We note that this derivation also holds for the case where ${\bf A}_{\rm ext}$
and ${\bf m}$ are lattice-periodic. In this case,
the surface term which had to be zero in order to derive Eq. (\ref{engyext}),
sums to zero over the faces of the periodic box.

\subsection{Showing $\widetilde{\bf B}_{\rm xc}$ is not the functional
 derivative of an $\widetilde{E}_{\rm xc}$}
If $\widetilde{\bf B}_{\rm xc}$ was indeed the functional derivative of a
scalar, then it would follow that the second derivative would be symmetric
in its arguments.
It is simple to demonstrate that
\begin{align*}
 \frac{\delta^2\widetilde{E}_{\rm xc}}{\delta{\bf m}({\bf r})\delta{\bf m}({\bf r}')}
 &\equiv\tilde{f}_{\rm xc}({\bf r},{\bf r}') \\
 &=f_{\rm xc}({\bf r},{\bf r}')
 +\int d^3 r''\,M({\bf r},{\bf r}'')f_{\rm xc}({\bf r}'',{\bf r}'),
\end{align*}
where
\begin{align*}
 M({\bf r},{\bf r}'')=\nabla_r\otimes\nabla_r\frac{1}{|{\bf r}-{\bf r}''|}
\end{align*}
and $f_{\rm xc}({\bf r},{\bf r}')$ is a $3\times 3$ matrix.
If $M$ had been a Dirac delta function then
$\tilde{f}_{\rm xc}$ would be symmetric in its arguments. This is not
the case though and hence $\widetilde{\bf B}_{\rm xc}$ is not the functional
derivative of an $\widetilde{E}_{\rm xc}$. However, $M$ is highly localized in
$|{\bf r}-{\bf r}''|$.

\subsection{Computational details}
The full potential linearized augmented plane wave (LAPW) method
implemented within the Elk
code\cite{elk} is used in the present work. All calculations are performed in
the presence of the spin-orbit coupling. 
To obtain the Pauli spinor states, the Hamiltonian containing only the scalar
potential is diagonalized in 
the LAPW basis: 
this is the first-variational step. The scalar states thus obtained are then
used as a basis to set 
up a second-variational Hamiltonian with spinor degrees of freedom\cite{singh}.
This is more efficient 
than simply using spinor LAPW functions, but care must be taken to ensure that
there is a sufficient number 
of first-variational eigenstates for convergence of the second-variational
problem. For example, $394$ states 
per {\bf k}-point were used for the pnictides to ensure convergence of the
second variational step. We use a {\bf k}-point set of 
$20 \times 20 \times 10$ for pinictides and $20 \times 20 \times 20$ for the
rest of the materials. A smearing width of 0.027eV was used.

\subsection{Separate scaling and removal of source term}

It is also interesting to investigate separately the effect of making the
functional source-free and
scaling it. We found that the purely source-free LSDA and GGA functionals lead to
highly under estimated 
moments (see Table \ref{tab_sm}). This is due to suppression
of the $z$-projected moment. The magnetization
density obtained by purely source-free LSDA and GGA is also highly
non-collinear (i.e. x and y projected moments 
are as significant as M$_{\rm z}$). The scaling alone of the LSDA/GGA has,
as expected, rather trivial effect of 
increasing the moment universally since the whole {\bf m}({\bf r}) is uniformly scaled.
One could envisage using the scaling
parameter so as to reproduce experimental moment for the materials. This,
however, has the disadvantage that the
scaling parameter is then highly material-dependent. The combination of the two things 
(scaling and source-free) leads to good agreement with experiments for
a wide set of materials and most
importantly the scaling factor depends only on the choice of original functional. 

\begin{table}[h]
\caption{Magnetic moment (in $\mu_{\rm B}$) per Fe atom for pnictides and per
 magnetic atom for the rest of the materials. LSDA$_s$/GGA$_s$ moments are calculated by enhancing
the LSDA, GGA magnetization density by 1.12 and 1.14 respectively. 
LSDA/GGA no source results are obtained by making LSDA/GGA source free (i.e. without any scaling)}\label{tab_sm}
\begin{tabular}{c|c|c|c|c|c}
Material       & Expt  & LSDA$_s$ & LSDA      & GGA$_s$ & GGA              \\ 
               &       &          & no source &         & no source         \\ \hline\hline 
LaFeAsO        & 0.63  & 2.11     & 0.07      & 2.42    & 0.06                 \\ \hline
CeFeAsO        & 0.8   & 2.12     & 0.02      & 2.37    & 0.01               \\ \hline
PrFeAsO        & 0.5   & 2.30     & 0.15      & 2.40    & 0.12              \\ \hline
NdFeAsO        & 0.54  & 2.20     & 0.32      & 2.43    & 0.30               \\ \hline
CaFe$_2$As$_2$ & 0.8   & 2.20     & 0.07      & 2.46    & 0.06               \\ \hline
SrFe$_2$As$_2$ & 0.94  & 2.24     & 0.06      & 2.50    & 0.04               \\ \hline
BaFe$_2$As$_2$ & 0.87  & 2.40     & 0.06      & 2.51    & 0.05               \\ \hline
FeTe           & 2.25  & 2.60     & 0.88      & 2.67    & 0.95              \\ \hline
bcc-Fe         & 2.2   & 2.62     & 1.91      & 2.66    & 1.90               \\ \hline
hcp-Co         & 1.7   & 1.82     & 1.28      & 1.87    & 1.35               \\ \hline
Ni             & 0.65  & 0.71     & 0.53      & 0.69    & 0.57               \\ \hline
Ni$_3$Al       & 0.077 & 0.26     & 0.01      & 0.29    & 0.06             \\ \hline
ZrZn$_2$       & 0.085 & 0.27     & 0.17      & 0.36    & 0.20            \\ \hline
\end{tabular}
\end{table}


\begin{thebibliography}{56}
\bibitem[{uj()}]{uj}
\bibinfo{note}{The value of $U$ and $J$ used with LSDA functional are
  $U=2.04$eV, $J=0.57$eV for NdFeAsO, $U=3.27$eV, $J=0.67$eV for PrFeAsO. The
  values used in combination of GGA are $U=4.7$eV, $J=1.4$eV for NdFeAsO,
  $U=2.04$eV, $J=1.15$eV for PrFeAsO.}
\end{thebibliography}

\end{document}